\documentstyle[aps,twocolumn]{revtex}

\begin{document}
\title
{Resistivity of the Two-Channel Kondo
Lattice Model
in Infinite Dimensions}

\author{D. L. Cox\\
Department of Physics, Ohio State University, Columbus, OH 43210}
\date{\today}
\maketitle
\setlength{\baselineskip}{12pt}

\widetext

\begin{abstract}
\smallskip
Analytic results for the resistivity of the two-channel Kondo lattice
in a particular infinite dimensional limit (Lorentzian density of
states)
are presented.  It is argued
that in the absence of symmetry breaking phase transitions or applied
fields
there is a residual resistivity at zero temperature
due to the spin disorder scattering off of the two-channel screening
clouds.  This may explain the unusual resistivity of UBe$_{13}$. For
the same limit
the single channel Kondo lattice is an insulator
states at particle hole symmetry and half filling, but metallic away
from particle-hole symmetry or in applied magnetic field.
\end{abstract}
\bigskip

PACS Nos.  74.70.Vy, 74.65.+n, 74.70.Tx
\narrowtext
The
heavy fermion materials such as UBe$_{13}$, UPt$_3$,  and
CeCu$_{2}$Si$_{2}$\cite{hfrevs} continue to resist comprehensive
theoretical
understanding.  In these intermetallic compounds,
strongly correlated 4f/5f orbitals on the rare earth or actinide sites
with localized magnetic moments give rise to anomalous properties
relative to normal metals. In particular, 100-1000 fold enhancement of
the linear coefficient in the electronic specific heat is observed,
giant resistivities (of order 100 $\mu-\Omega$-cm) which are
non-monotonic in temperature are typical.  The three materials above
(and
several others) exhibit superconducting transitions where the heavy
electrons pair.

The
simplest model which has been used to understand these complex
materials is the Kondo lattice Hamiltonian, given by
\begin{equation}
H_1 = \sum_{\vec k,\sigma} \epsilon_k c^{\dagger}_{\vec k\sigma}
c_{\vec k,\sigma} - {{\cal J}\over N_s} \sum_{\vec R} \vec S(\vec
R)\cdot
\vec s_c(\vec R)
\label{ham}
\end{equation}
where $c_{\vec k\sigma}$ destroys a spin $\sigma$ conduction electron
of
momentum $\vec k$ and energy $\epsilon_k$, and the conduction spins
interact
antiferromagnetically (${\cal J}<0$) with the $S=1/2$ local moments
situated
at sites $\{\vec R\}$.  Here $\vec S(\vec R)$ is a local moment spin
operator,
and $\vec s_c(\vec R)$ a conduction spin operator.

This model has been extensively studied in
one dimension\cite{onedrefs}; the related symmetric Anderson model has
been studied in
infinite dimensions\cite{infdrefs}.  From these works it is clear that
at half filling, for sufficiently large ${\cal J}$, this model should
have an
insulating ground state
which is readily seen from counting arguments.  Away from half filling,
the model is
metallic
in the
absence of any symmetry breaking effects.

By comparison, essentially nothing
is known about the two-channel Kondo lattice
model, in which
two degenerate species of itinerant electrons interact with a lattice
of local spin 1/2
moments.  This model
is of considerable interest since in the impurity limit  (single local
moment) it contains a
non-trivial
ground state which is critical (possessing infinite range spatial
correlations)
and a non-Fermi liquid excitation
spectrum\cite{nozbland,ludaffgrn,aflupaco,emkiv}.
The Hamiltonian
\newpage\hfill\\*[2.0cm]
\smallskip
is obtained by adding a channel index $\alpha=\pm$ to the conduction
creation, annihilation, and spin operators in  Eq. (\ref{ham}).  In
applying this model to describe the heavy fermion materials,
the channel index will either be a local orbital index for the magnetic
Kondo
effect, or a local spin index for the quadrupolar Kondo effect.  See
Ref.
\cite{coxsend} for further details.

Impetus to study the two-channel lattice model
is provided by: (i) some properties of
heavy electron alloys are described well by
two channel Kondo models--Y$_{1-x}$U$_x$Pd$_3$\cite{yupd},
Th$_{1-x}$U$_x$Ru$_2$Si$_2$,\cite{thursi}, and
La$_{1-x}$Ce$_x$Cu$_{2.2}$Si$_2$\cite{lacecusi}.
(ii) UBe$_{13}$ has been proposed as a two-channel
Kondo lattice material\cite{coxold}. It also has an extremely unusual
resistivity $\rho(T)$:   at
the superconducting transition temperature $T_c$, $\rho(T_c)$
is extremely large (order 100 $\mu\Omega$-cm, close to the unitarity
limit
in which every atom scatters resonantly) and is rapidly
suppressed in applied magnetic field\cite{batold,andnew} {\it and}
applied
pressure\cite{aronson}.  The latter fact makes it unlikely that the
resistivity is
due to any ordinary dirt.

In this paper, new results are presented for the form of
resistivity of the two-channel Kondo lattice model in a certain
infinite
dimensional limit (Lorentzian bare density of states [DOS] for
itinerant electrons).
For the paramagnetic state that the
lattice behaves as an
incoherent metal, with finite residual resistivity in the absence of
applied
or spontaneous symmetry breaking fields.
Application
of a magnetic field or channel spin field induces a cross-over to a
Fermi liquid at low temperatures, below which the resistivity should be
described by a
universal scaling function.   Since the Lorentzian
DOS is pathological in possessing only one non-vanishing
moment,  I control these results by demonstrating that
reasonable physics emerges in applying the same method to the
one-channel Kondo lattice.  In that case the model gives (in the
absence of applied or spontaneous symmetry breaking
fields) an insulating
ground state at half filling and
a metallic ground state
away from half-filling as anticipated from general considerations and
from
studies of the Anderson
lattice model in infinite  dimensions\cite{infdrefs}.

All results follow from the asymptotic low energy, low temperature
forms
of
the one particle $T$-matrices for scattering off one
spin 1/2 impurity by either one or two channels of conduction
electrons.
The infinite dimension limit is then obtained by self-consistently
embedding the impurity in a manner prescribed below.  I normalize the
$T$-matrices by multiplying
by the number of sites.

For the one-channel Kondo model, at particle-hole
symmetry, the zero temperature retarded one-particle $T$-matrix,
denoted as $t(\omega,T)$ is given by
\begin{eqnarray}
t(\omega,T=0)& \approx& -{i\over \pi N(0)} \sin^2({\pi\over 2})
\nonumber \\
&&[1 +i
a{\omega\over k_BT_K}
-b({\omega
\over k_B T_K})^2 + ....]
\label{tone}
\end{eqnarray}
where $N(0)$ is the Fermi level density of states, $T_K$ is the Kondo
temperature, $a$ is a (universal) pure number whose value
is unimportant for our purposes, and $b=0.104$\cite{hewsonplus}.  The
Kondo
scale is related to the exchange coupling ${\cal J}$ according to
$k_BT_K \approx D(N(0){\cal J})^{1/2}\exp(1/N(0){\cal J})$ where $D$ is
the conduction bandwidth.  The scattering strength at the Fermi energy
for the
above $T$-matrix is at the unitarity limit with a maximal,
resonant phase
shift of $\delta =\pi/2$; the effective single particle picture at
$T=0$ follows from the singlet formation below $T_K$
which lifts the spin degeneracy.  The linear energy dependence in
$Re{t}$
describes the large effective mass imparted to the conduction electrons
in the vicinity of
the impurity.
The $\omega^2$ behavior of the next leading
imaginary contribution follows from the Fermi liquid behavior of the
quasi-particle
states surrounding the impurity. At finite temperature, a
term proportional
to $T^2$ is to be added to this next leading contribution.  Note that
particle-hole symmetry gives the special result that the $T$-matrix is
purely imaginary at the Fermi energy.

For the two-channel Kondo model at particle-hole symmetry, the zero
temperature retarded one-particle $T$-matrix is given
by\cite{ludaffgrn,coxruck}
\begin{equation}
t(\omega,T=0) \approx -{i\over2 \pi N(0)}[1
- \tilde a (1
+ i sgn(\omega))\sqrt{{|\omega|\over k_BT_K}}
+ ....]
\label{ttwo}
\end{equation}
where $\tilde a$ is a universal, pure pure number whose value is
unimportant for our purpose.  Note that the next leading terms
are non-analytic, which derives from the non-Fermi liquid character of
the
excitation spectrum.
At finite temperature, $\sqrt{T}$ contributions will arise in the above
expression.
The crucial features of Eq. (\ref{ttwo}) in rendering the two-channel
lattice a metal are (i) the factor of 1/2 reduction from the unitarity
limit in the Fermi energy value, and (ii) the fact that the T-matrix is
purely imaginary at the Fermi energy, in the absence of symmetry
breaking effects.

To proceed to the lattice, a relation of the $T$-matrix to the local
self
energy $\Sigma(\omega,T)$ is needed.  Using the defining relation for
the $T$-matrix and Dyson's
equation  gives
\begin{equation}
\Sigma(\omega,T) = {t(\omega,T) \over 1 + G_0(\omega) t(\omega,T)}
\label{sigma}
\end{equation}
where $G_0(\omega)$ is the non-interacting on-site Green's function.

 The infinite
dimensional limit renders the
conduction electron self energy purely local\cite{infdreview}.
This simplifies the problem greatly: one must solve self consistently
the ``impurity'' problem for one site removed from the lattice in which
all other sites feel the
impurity self energy corresponding to the removed site.
The self-consistency is implicit in the equation
relating the local electronic
propagator $G(\omega,T)$
to the momentum space one, which is
\begin{equation}
G(\omega,T) = {1\over N_s} \sum_{\vec k} {1\over \omega -
\epsilon_{\vec k} - \Sigma(\omega,T)}
\label{glocal}
\end{equation}
where the electrons have single particle energy $\epsilon_{\vec k}$
measured with respect to the Fermi energy
and live on an $N_s$-site lattice.

The Lorentzian DOS corresponds to a peculiar limiting
large $d$-lattice with
infinite range oscillatory hopping along each of the hypercubic
principal directions\cite{sirefs}.
The simplifying feature of the Lorentzian DOS is that
self consistency in Eq. (\ref{glocal})
is automatic\cite{sirefs}.  To see this, first take the non-interacting
density of states
$N_{0}(\epsilon) = {D / \pi(\epsilon^2+D^2)}$
to obtain $G_0(\omega) = 1/(\omega + iD)$.  Converting sum to
integral
in Eq. (\ref{glocal}) gives
$G(\omega,T) = 1/(\omega + iD - \Sigma(\omega,T))$ which clearly
satisfies Dyson's equation.
Hence, solving the problem on the lattice for a Lorentzian DOS means
you can just
plug in the impurity results for the corresponding DOS
provided there is no significant shift of the electronic chemical
potential\cite{sirefs}.  For particle-hole symmetry,
this last point is safely ensured.

For the one channel model with the Lorentzian bare DOS,
it follows that
the resulting renormalized one particle DOS in the presence of the
interactions
vanishes quadratically
in $\omega,T$ for low frequency,temperature in the case of particle
hole symmetry.  Explicitly,
\begin{equation}
N(\omega) = -{1\over \pi} ImG(\omega+i0^+,T)
\approx {b \over \pi D}(\omega^2 + \pi^2 T^2)
\label{dosone}
\end{equation}
As a corollary, the resistivity diverges like $T^{-2}$.  The reason for
this is two fold.  First,
the imaginary part of the self energy may be seen by similar analysis
to that of Eq. (\ref{dosone}) to diverge
as $-Im\Sigma(0,T) \approx D/b \pi^2T^2$.  Second, there are no
conductivity vertex corrections in infinite
dimension due to the locality of the interacting
vertex\cite{mullhart}.  Hence, $\rho(T) \sim
1/<Im\Sigma(\omega,T)>_{FS}$
where the angular brackets denote the usual Fermi surface transport
average\cite{mahan}.  Thus, at particle-hole
symmetry the single channel Kondo lattice is an insulator
for this Lorentzian bare DOS with a
``soft gap'' (power law vanishing of the density of states).

Modifying the Fermi level phase shift for the single channel case to
 $\delta\ne \pi/2$
allows us to go away from
particle hole symmetry.  The real part of the Fermi level self energy
is
then
given by
$\Sigma(0,0) = D\tan(\delta)$, while the imaginary part
$\Sigma(\omega,T)$ now
vanishes as $\omega^2+\pi^2T^2$.
Thus,
away from particle hole symmetry the Lorentzian bare DOS gives a
metallic behavior
at the Fermi energy.
This also simulates the effect of doping which
tunes the available screening charge from unity and thus modifies the
phase
shift according to
the Friedel sum rule.
The application of a magnetic field $H$ will
modify the $\pi/2$ zero field phase shift in the $\pm$ spin channel to
$\pm[\pi/2 - c
(\mu_{eff} H/k_BT_K)^2 + ...]$
for low fields, and thus will produce a continuous metal-insulator
transition for this Lorentzian bare DOS.  In this case,
the height of the Fermi level renormalized DOS will grow quadratically
in the
applied field strength.
As in previous average $T$-matrix calculations (which are exact in
infinite dimensions for a Lorentzian bare DOS )\cite{coxgrewe},
I anticipate that the high temperature behavior of the resistivity
will be little different from the impurity limit.
These results for a non-particle-hole symmetric scattering and applied
magnetic field agree qualitatively with previous calculations for Kondo
lattice materials\cite{onedrefs,infdrefs,bedellref}.  Differences
in
detail are due to the anomalous tails of the Lorentzian DOS.

In the case of the two-channel model, using Eq. (\ref{ttwo})
it is straightforward to see that the self energy at
zero temperature and low frequency goes as
$-Im\Sigma(\omega,0) \approx D[1-2\tilde a\sqrt{|\omega|/ k_BT_K}
+ ...]$
so that the resistivity at zero temperature will be at the
unitarity limit
(scattering rate $1/\tau = 2D/\hbar$) and the lattice
is an incoherent metal in the absence of any symmetry breaking.
Concommitantly, the
one electron density of states at the Fermi energy will exhibit a
square root cusp at $E_F$.

The physical interpretation of this remarkable result is
straightforward:  in the two channel
case the degeneracy of the impurity spin is never lifted.  Hence in the
absence of a spin ordering transition which lifts the degeneracy, or a
superconducting
transition,
each
cloud contributes
a ``spin disorder scattering'' which leads to a `violation' of
Bloch's theorem\cite{ludaffgrn,coxruck,gancole}.
 Such a result is reminiscent of magnetic rare earth
intermetallics such as Gd\cite{rebook},  but such resistivities are
well
understood through lowest order golden rule estimates of the scattering
from the bare local moments, rather than a complex many body spin
cloud.

The application of a spin or channel symmetry breaking field
for a single
two-channel impurity drives the physics to that of a Fermi
liquid\cite{aflupaco}.
For the case of a field $H_{sp}$ which couples linearly
to the impurity spin,
it is well established that the Fermi level one-particle phase shift
tends to $\pm \pi/4$ for
spin $\pm$\cite{aflupaco}.  The
crossover temperature for this behavior is $T_{sp}= H_{sp}^2/T_K$ (the
moment is included in $H_{sp}$).
Hence, for $T>T_{sp}$,
the system looks like a two-channel lattice
with $\rho(T)$
headed
towards the unitarity limit at low $T$.  For $T<T_{sp}$,
$\rho(T)$ must
drop towards zero with a $T^2$ Fermi liquid behavior.  For an applied
channel field $H_{ch}$ coupling linearly to the channel spin,
one channel will pop to zero phase shift at $T=0$, the other to
$\pi/2$, making for
a ``half insulator/half metal'':  the zero phase shift channel will
short circuit the resonant
channel, leading to zero resistance at $T=0$, but no conduction will
be possible in the
resonant channel.   The crossover temperature in this case is
$T_{ch}=H_{ch}^2/k_BT_K$. In either case, the resistivity
should be described by a universal scaling function below the crossover
scale (based on the results of Ref. \cite{ludaffgrn}), of the form
$\rho(H_{\alpha},T) \sim F(H_{\alpha}^2/T)$ where
$\alpha=sp,ch$.  The schematic resistivity behavior expected for the
two-channel model
is illustrated in the Figure.

These resistivity results are obtained within an
enforced normal, paramagnetic phase.  The onset of a magnetic,
superconducting, or channel field instability would clearly
alter the structure of the schematic resistivity curves.

I now turn to a discussion of the heavy fermion materials and the
possible relevance of
these considerations there.  For UBe$_{13}$, it has been proposed that
a two-channel quadrupolar Kondo lattice model may provide an
appropriate description\cite{coxold}.  This metal has,
reproducibly, $\rho(T_c) \simeq 100\mu-\Omega-cm$. The extrappolated
value for $\rho(0)$ is nearly as large  and (i) vanishes in applied
pressure, (ii) goes
to zero in applied field with a scaling function form $\rho(H,T) \sim
F(H^{\beta}/(T-T_0))$ where
values of $\beta=1.0,T_0=0$\cite{batold} and
$\beta=1.67,T_0=0.75K$\cite{andnew} have been observed.
 Since there is no
evidence of magnetic or quadrupolar order at any temperature above the
superconducting
transition, I speculate that the unusual residual
resistivity of this material may   provide an example of the incoherent
metal scenario describe above, with the differing
values of $\beta,T_0$ from the infinite dimension Lorentzian DOS
description
deriving from finite dimensionality effects.  In
this case the magnetic field should initially behave as a channel
field,
since the effective impurity spin is quadrupolar and the channel index
is magnetic, while for sufficiently large fields the magnetic field
induced quadratic splitting of the non-magnetic ground state will
render it an effective spin field.

For UPt$_3$, a quadrupolar Kondo lattice Hamiltonian may also prove a
suitable starting point
for theoretical modeling\cite{coxsend}, although the low temperature
behavior in this system is clearly that
of a Fermi liquid and weak in-plane magnetic order is observed
in this material\cite{hfrevs}.  The in-plane order serves as a
channel symmetry breaking field.   Since the crossover effects in the
two-channel model scale with the square of the applied field, I
speculate that it is sufficient to have a molecular field induced value
to $H_{ch}^2$ which
is certain to be non-vanishing for a lattice model.  Thus the Fermi
liquid scale would be set
by the crossover temperature proportional to the molecular field
induced value of $H_{ch}^2/T_K$.   To study this idea further dilution
on the uranium sublattice is
desireable; unfortunately no suitable reference compound exists.
URu$_2$Si$_2$ offers a more promising possibility,
particularly in view of evidence for the
two channel Kondo effect in Th$_{1-x}$U$_x$Ru$_2$Si$_2$\cite{thursi}.

In summary, a study has been made of the low temperature resistivity of
the one and two-channel Kondo lattice models in infinite dimension
assuming an underlying Lorentzian DOS for the conduction
electrons in an enforced paramagnetic state.  At particle hole
symmetry, the two
channel lattice is an incoherent metal, with finite residual
resistivity due to spin-disorder
scattering off of the degenerate two-channel screening clouds. This may
be altered to ordinary metallic behavior by
application of spin or channel symmetry breaking fields.  This novel
result may explain the
unusual resistivity of UBe$_{13}$.  Though the Lorentzian DOS gives
reasonable results for the one channel model (e.g., an insulator at
half filling), it will be important to
study a non-Lorentzian DOS in the future to ensure that the results
obtained are not pathological.

It is a pleasure to acknowledge useful conversations with M. Aronson,
M. Jarrell, A. Ludwig, H. Pang, Th. Pruschke, A. Ruckenstein, and J.W.
Wilkins.  This research was supported by a grant
from the U.S. Department of Energy, Office of Basic Energy Sciences,
Division of Materials
Research.

{\bf Figure }. Schematic form of the resistivity of the two-channel
Kondo lattice in infinite dimensions for a bare Lorentzian conduction
density of states.  The solid line is for no symmetry breaking, the
dashed line
in the case of applied spin or channel field $H_{sp,ch}$, with
$T_{sp,ch} = H_{sp,ch}^2/T_K$.

\end{document}